\documentclass[notitlepage,aps,epsfig,showpacs,floats,twocolumn,amssymb,amsmath,floatfix,
groupedaddress,superscriptaddress]{revtex4-2}
\usepackage{graphicx} 
\usepackage{subfigure}
\usepackage{dcolumn} 
\usepackage{bm} 
\usepackage{float} 
\usepackage{bbm} 
\usepackage{enumitem,xcolor}
\usepackage{color}
\usepackage{amsfonts}
\usepackage{xcolor}
\usepackage{lmodern} 
\usepackage{natbib}
\usepackage{comment}
\usepackage{amsmath} 
\usepackage{amsmath,amssymb}
\usepackage{natbib}
\usepackage[margin=2cm]{geometry}
\usepackage{tikz}
\usepackage{ulem}
\usetikzlibrary{matrix}

\newcommand\bi{\begin{itemize}}
\newcommand\ei{\end{itemize}}

\newcommand{\adDeriv}[1]{\frac{\mathrm{D}#1}{\mathrm{D}t}} 
\DeclareMathOperator{\reynolds}{\mathrm{Re}}
\DeclareMathOperator{\SO}{\mathrm{SO}}

\usepackage[colorlinks=true,citecolor=blue]{hyperref}
\hypersetup{colorlinks=true,citecolor=blue,linkcolor=blue,urlcolor=black}

    \usepackage{geometry}

    \usepackage{tikz} 
    \usetikzlibrary{shapes,arrows,calc}
    \usetikzlibrary{datavisualization}
    \usetikzlibrary{datavisualization.formats.functions}
    \usepackage{pgfplots} 
    \usepackage{tikz}
\usetikzlibrary{matrix}  

\usepackage[colorlinks=true,citecolor=blue]{hyperref}

\begin{document}

\title{Symmetry-restoring crossover from defect-free to defect-laden turbulence in polar active matter}
\author{Benjamin H. Andersen}
\affiliation{Niels Bohr Institute, University of Copenhagen, Blegdamsvej 17, Copenhagen, Denmark}
\author{Julian Renaud}
\affiliation{École Normale Supérieure, PSL Research University, 45 rue d’Ulm, 75005 Paris, France}
\author{Jonas R\o nning}
\affiliation{Njord Centre, Department of Physics, University of Oslo, P. O. Box 1048, 0316 Oslo, Norway}
\author{Luiza Angheluta}
\email{luiza.angheluta@fys.uio.no}
\affiliation{Njord Centre, Department of Physics, University of Oslo, P. O. Box 1048, 0316 Oslo, Norway}
\author{Amin Doostmohammadi}
\email{doostmohammadi@nbi.ku.dk}
\affiliation{Niels Bohr Institute, University of Copenhagen, Blegdamsvej 17, Copenhagen, Denmark}

\begin{abstract}
Coherent flows of self-propelled particles are characterized by vortices and jets that sustain chaotic flows, referred to as active turbulence.  Here, we reveal a crossover between defect-free active turbulence and active turbulence laden with topological defects. Interestingly, we show that concurrent to the crossover from defect-free to defect-laden active turbulence is the restoration of the previously broken $\SO(2)$-symmetry signaled by the fast decay of the two-point correlations. By stability analyses of the topological charge density field, we provide theoretical insights on the criterion for the crossover to the defect-laden active turbulent state. Despite the distinct symmetry features between these two active turbulence regimes, the flow fluctuations exhibit universal statistical scaling behaviors at large scales, while the spectrum of polarity fluctuations decays exponentially at small length scales compared to the active energy injection length. These findings reveal a new dynamical crossover between distinct spatiotemporal organization patterns in polar active mater.
\end{abstract}
\maketitle

\section{Introduction}
Recent developments in active matter have shed light on a form of chaotic flows sustained by living microswimmers at low Reynolds numbers, often known as {\it active turbulence}~\cite{alert2021active}. A distinctive feature of active turbulence, compared to classic inertial turbulence, is the continuous injection of energy at the scale of individual constituents of the active material. Striking examples are dense bacterial suspensions~\cite{Wensink2012,meacock2021bacteria}, cellular monolayers~\cite{blanch2018turbulent,lin2021energetics}, and assemblies of subcellular filaments~\cite{Sanchez2012,martinez2021scaling}, that are all composed of active elements - i.e., individual bacterium, single cell, or motor proteins walking on single filaments - each capable of converting chemical energy to mechanical work~\cite{Marchetti13,Bechinger2016,Doost18}.
Although, most of the work so far has been limited to visual resemblance with inertial turbulence, recent works suggest the existence of universal scaling features in active turbulence~\cite{alert2020universal,martinez2021scaling} and even show~\cite{doostmohammadi2017onset} that crossover to active turbulence belongs to the same directed percolation universality class as in the inertial turbulence~\cite{lemoult2016directed}.

Nevertheless, a majority of studies of active turbulence so far, including the works suggesting universal scaling laws and universality classes of turbulence transition, have focused on a subclass of active materials known as {\it active nematics}~\cite{ngo2014large,urzay2017multi,krajnik2020spectral,carenza2020cascade,coelho2020propagation,chandragiri2020flow}, which models flows generated by dense assemblies of elongated particles~\cite{doostmohammadi2018active}. Within the active nematic framework, the particles are essentially characterized as shakers~\cite{ramaswamy2010mechanics}: they neither have any polarity or ability to propel themselves, and instead generate head-tail symmetric (nematic) active stresses in the fluid. The active stresses in turn drive hydrodynamic instabilities in the flow and the orientation field of particles, resulting in a chaotic flow state that is interleaved with topological defects - singular points in the orientation field where the order breaks down~\cite{Marchetti13,Doost18,ardavseva2022topological}.

Previous studies of polar active matter have characterized spontaneous flows~\cite{voituriez2006generic,giomi2008complex}, and associated nonequilibrium steady-states~\cite{tjhung2011nonequilibrium,giomi2012polar}.
On the other hand, studies of self-propulsion effects on active turbulence have mostly neglected orientational couplings. Instead, the focus was put on generalized Navier-Stokes equations, where activity is introduced through addition of phenomenological higher-order terms in the momentum equation, corresponding to a second order negative viscosity and a fourth-order hyper-viscosity, to give characteristic vortex length to active turbulence~\cite{Wensink2012,bratanov2015new,linkmann2019phase,linkmann2020condensate}. As such, characteristics of active turbulence in models of polar active fluids in the presence of topological defects were only marginally explored in the context of polar active emulsions~\cite{carenza2020multiscale} and polar flocks with inertia~\cite{chatterjee2021inertia}. Additionally, notwithstanding the interesting recent characterization of the annihilation of topological charges in colloidal flocks~\cite{chardac2021topology} and despite the emerging roles of topological defects in various biological fluids~\cite{doostmohammadi2021physics,guillamat2022integer,shankar2022topological,endresen2021topological}, the majority of research so far has focused on nematic topological defects~\cite{Doost18,shankar2022topological,vafa2020multi} and studies of active turbulence in polar fluids in the presence of polar topological defects are lacking.
\begin{figure*}[th]
\centering
  \includegraphics[width=0.9\linewidth]{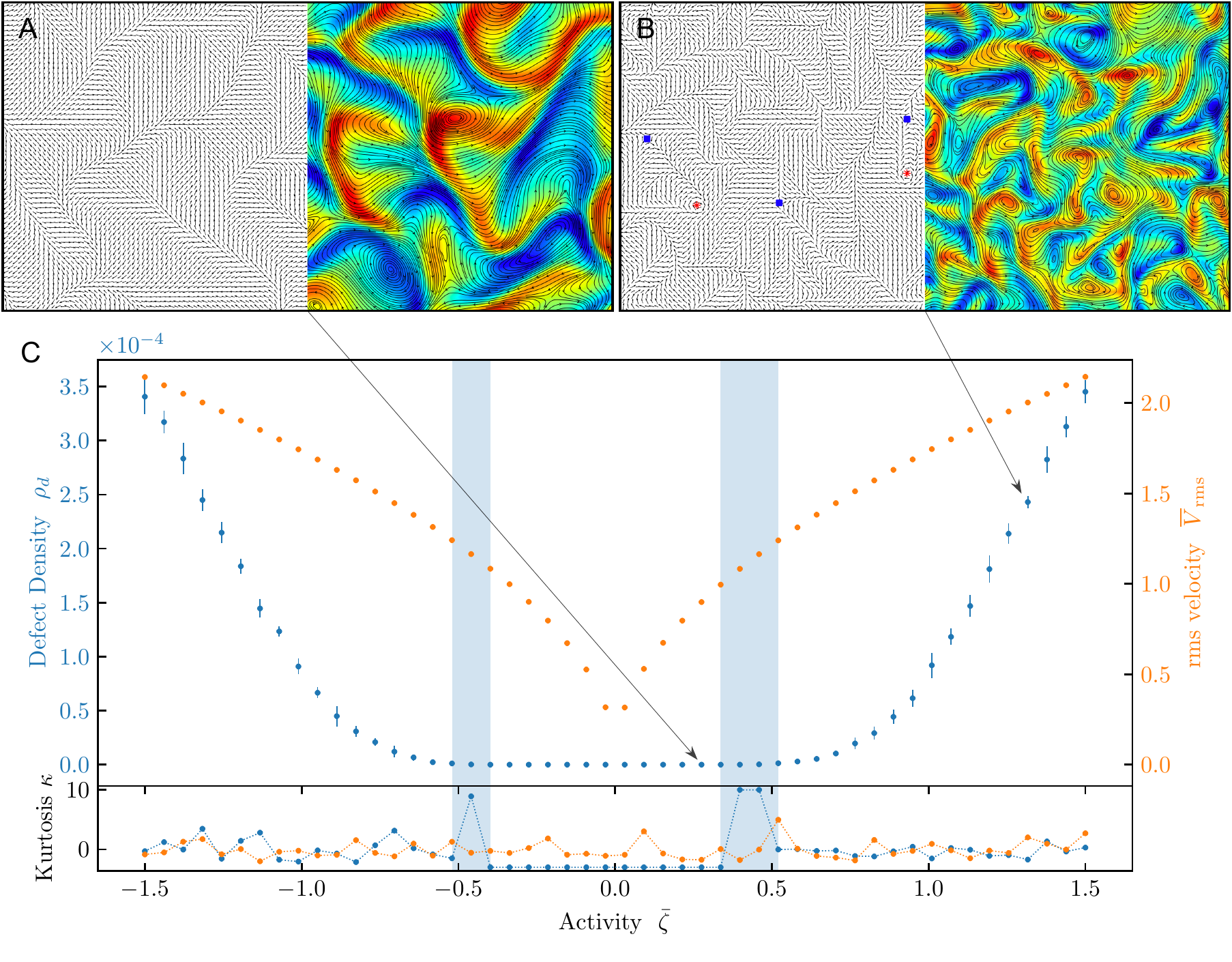}
  \caption{{\bf Crossover from defect-free to defect-laden active turbulence upon increasing strength of active stresses.} Shown here are the averaged defect density (blue dots) and the rms-velocity (orange dots) in units of $V_p = \sqrt{AK}/\eta$, characterizing the passive relaxation velocity of the polar particles. Active turbulence arises as soon as the activity is non-zero, but topological defects form and persist in the polarity field only beyond a finite threshold both in contractile ($\zeta < 0$) and extensile systems ($\zeta > 0$).  Results were averaged over time and for 10 simulations from different initial conditions, the error bars standing for the corresponding standard deviation.
  The excess kurtosis $\kappa_i$ of both quantities are also represented at the bottom: for large activity levels, $\kappa(\rho_d) \simeq 0$, which indicates the population of defects is typically equilibrium-like and follows a normal distribution law. For small activity levels, there are no defects at all and $\kappa$ is set by convention to $-3$. In between, the crossover is characterized by very rare formation and annihilation events, as evidenced by the peaks in $\kappa$ (blue highlighted areas). There are no such peaks for the kurtosis of the rms-velocity, which indicates there is no equivalent crossover regime. Insets illustrate polarity field ({\it left panels}) and the corresponding velocity field ({\it right panels}) characterized by streamlines (solid black lines) colored by the normalized vorticity $\omega/\omega_{\text{max}}$. Only a small subset of the full domain ($1/16th$ zoomed-in region) is shown. Within the polarity field positive and negative integer topological defects are marked by red asters and blue squares, respectively.}
  \label{fig:den}
\end{figure*}

Here we use numerical simulations of continuum polar active matter together with analytical arguments to shed light on the turbulence characteristics of active fluids laden with topological defects, accounting for both hydrodynamic effects and polar ordering. We first show how increasing active stresses results in a regime shift from defect-free active turbulence to active topological turbulence, where chaotic flows are interleaved with chaotic arrangements of full-integer topological defects. We provide theoretical arguments to predict the activity threshold for the crossover to active topological turbulence, based on the stability analyses of the topological charge density field. We further show that ordering and symmetry characteristics are different within these two dynamical regimes: while $\SO(2)$-symmetry is broken within the defect-free active turbulence, the defect-laden turbulence restores the global $\SO(2)$-symmetry. Additionally, we demonstrate the existence of universal scaling behavior in the power spectra of enstrophy and polarity. This is observed for all activities and within both \textit{defect-free} and \textit{defect-laden} active turbulence. 

\section{Methods}\label{sec:meth}
\subsection{Hydrodynamic model}
We consider an incompressible polar active fluid in two spatial dimensions, with the local orientational order described by a two-component order parameter $\mathbf{p}$ pointing in the direction of self-propulsion. Let then $\mathbf{u}$ and $\rho$ be the velocity and density of the polar fluid, respectively. Coarse-grained hydrodynamic equations can be derived by phenomenological considerations~\citep{Marchetti13,chatterjee2021inertia,Amiri_2022} and take the form 
\begin{subequations}
    \begin{align}
        &\adDeriv{u_i} = \frac{1}{\rho}\partial_j\sigma_{ij}, \label{eq:u}\\
        &\adDeriv{p_i} - \left(\lambda E_{ij} + \Omega_{ij}\right)p_j = \frac{1}{\gamma}h_i, \label{eq:p}
    \end{align}
\end{subequations}
along with the incompressibility condition $\partial_i u_i = 0$. Here $\adDeriv{} = \partial_t + u_j\partial_j$ is the usual advective derivative. The strain rate tensor $E_{ij}= (\partial_i u_j + \partial_j u_i)/2$ and the vorticity tensor $\Omega_{ij} = (\partial_i u_j - \partial_j u_i)/2$ are respectively the symmetric and anti-symmetric parts of the velocity gradient tensor. In the momentum balance equation~\eqref{eq:u}, the stress tensor $\sigma_{ij}$ is broken into a sum of three parts; viscous $\sigma_{ij}^\text{vis.} = 2\eta E_{ij}$, passive $\sigma_{ij}^\text{pas.} = -P\delta_{ij} + C_{ijkl}p_kh_l$, and active stresses $\sigma_{ij}^\text{act.} = -\zeta\left(p_ip_j - \frac{p^2}{2}\delta_{ij}\right)$~\cite{prost2015active,julicher2018hydrodynamic}. The first term in the passive stress is the usual hydrodynamic pressure. The second term accounts for elastic stresses through the anisotropic tensor $C_{ijkl} = \frac{\lambda+1}{2}\delta_{ik}\delta_{jl} - \frac{\lambda}{2}\delta_{ij}\delta_{kl} + \frac{\lambda-1}{2}\delta_{il}\delta_{jk}$, with $\lambda$ the flow alignment parameter and 
$\mathbf{h}=-\delta\mathcal{F}/\delta \mathbf{p}$~\cite{de1993physics,chandragiri2019active} the molecular field defined here from the free energy 

\begin{equation}
     \mathcal{F}=\int\mathrm{d}^2\mathbf{x}\left\{A\left(-\frac{|\vec p|^2}{2}+\frac{|\vec p|^4}{4}\right)
     +\frac{K}{2}\partial_ip_j\partial_ip_j \right\}.\label{eq:F}
\end{equation}
The free energy contains a local energy density with an energy scale $A$ that controls the isotropic-polar transition favoring the emergence of finite polarity at $|\mathbf{p}|=1$, and a non-local energy contribution with an elastic constant $K$ that penalizes deformations in the polarity field~\cite{frank1958liquid}. It is important to note that we do not treat the polarity as a unit vector, with fixed magnitude. Instead, in this formulation the polarity modulus is an important dynamical variable.
\subsection{Numerical method and the simulation parameters}
We simulate equations~(\ref{eq:u},\ref{eq:p}) using a hybrid lattice-Boltzmann method, combining finite-difference method for the evolution of polarity vector Eq.~\eqref{eq:p}, and the lattice-Boltzmann method for solving the incompressible Navier-Stokes equation Eq.~\eqref{eq:u} with $\rho=40$ and $\eta=3.6$ in lattice Boltzmann units, ensuring that the Reynolds number in the simulations is negligible ($\reynolds \ll 1$)~\cite{thampi2014instabilities,doostmohammadi2017onset}. The other relevant dimensionless numbers describing the system are: (i) the  dimensionless ratio of the viscosities $\eta/\gamma$, (ii) ratio of micro to macro length scales $(\sqrt{K/A})/$L, (iii) the flow alignment parameter $\lambda$, and (iv) the dimensionless active stress $\bar{\zeta}=\zeta/A$. Unless otherwise stated, we fix the viscosity ratio to $\eta/\gamma=3.6$, micro to macro length scale to $(\sqrt{K/A})/L=2\times 10^{-4}$ (assuring that the coherence length $l_p = \sqrt{K/A}$ is significantly smaller than the domain size $L$), and the flow alignment parameter to $\lambda=0.1$

Simulations were initialized with quiescent velocity field and noisy polar alignments close to the uniformly oriented state $\mathbf{p} = \hat{\mathbf{e}}_x$ under periodic boundary conditions, on quadratic domains of linear dimension $L=1024$, unless otherwise is stated.

\section{Results}
\subsection{Activity-induced crossover to defect-laden turbulence}

We begin by introducing global measures of the flow fluctuations and polar order parameter as functions of the activity parameter $\bar{\zeta}$. The flow is globally characterized by the dimensionless root mean squared (rms) velocity $\overline{V}_\text{rms}=V_\text{rms}/V_p$ that is normalized by the characteristic velocity of passive relaxation of polarity $V_p = \sqrt{AK}/\eta$. The polar order parameter is associated with the SO(2) symmetry and carries full-integer topological defects. Its global measure is given by the average density of topological defects $\rho_d = \langle N_d\rangle /L^2$. The averaging is done both over space and time in the statistical steady-state regime. 
As evident from Fig.~\ref{fig:den}, increasing activity beyond a certain threshold results in a continuous increase in the defect density for both extensile ($\zeta > 0$) and contractile ($\zeta < 0$) systems. We have carefully checked that the activity thresholds do not depend on the system size $L$, by simulating different domain sizes of $512 \times 512$, $1024 \times 1024$, and $4096 \times 4096$ and finding precisely the same quantitative dependence of defect density on the activity (see Fig. 2; {\it square and circle symbols)}. Interestingly, simultaneous measurement of the rms velocity $\overline{V}_\text{rms}$ indicates that even below the critical activity for defect nucleation, active stresses disturb the polarity field such that spontaneous flows are generated within the system, for any non-zero activity (Fig.~\ref{fig:den}C). 
A closer look at the velocity field of the system below the defect nucleation threshold shows that chaotic flows characterized by flow vortices and jets span the system (Fig.~\ref{fig:den}A; only a fraction of the entire domain is shown), although no topological defects are present in the polarity field. Qualitatively similar chaotic flows manifest at higher activities with a smaller typical length scale $L_\text{act.}$ defined below, though with the distinctive feature that the flows are now laden with the presence of full-integer topological defects (unbound vortex and antivortex pairs) in the polarity field (Fig.~\ref{fig:den}B). Together, these results establish that there is a well-defined activity threshold for the crossover from defect-free and defect-laden active turbulence in polar active matter.
\begin{figure}[t!]
    \centering
    \includegraphics[width=1.0\linewidth]{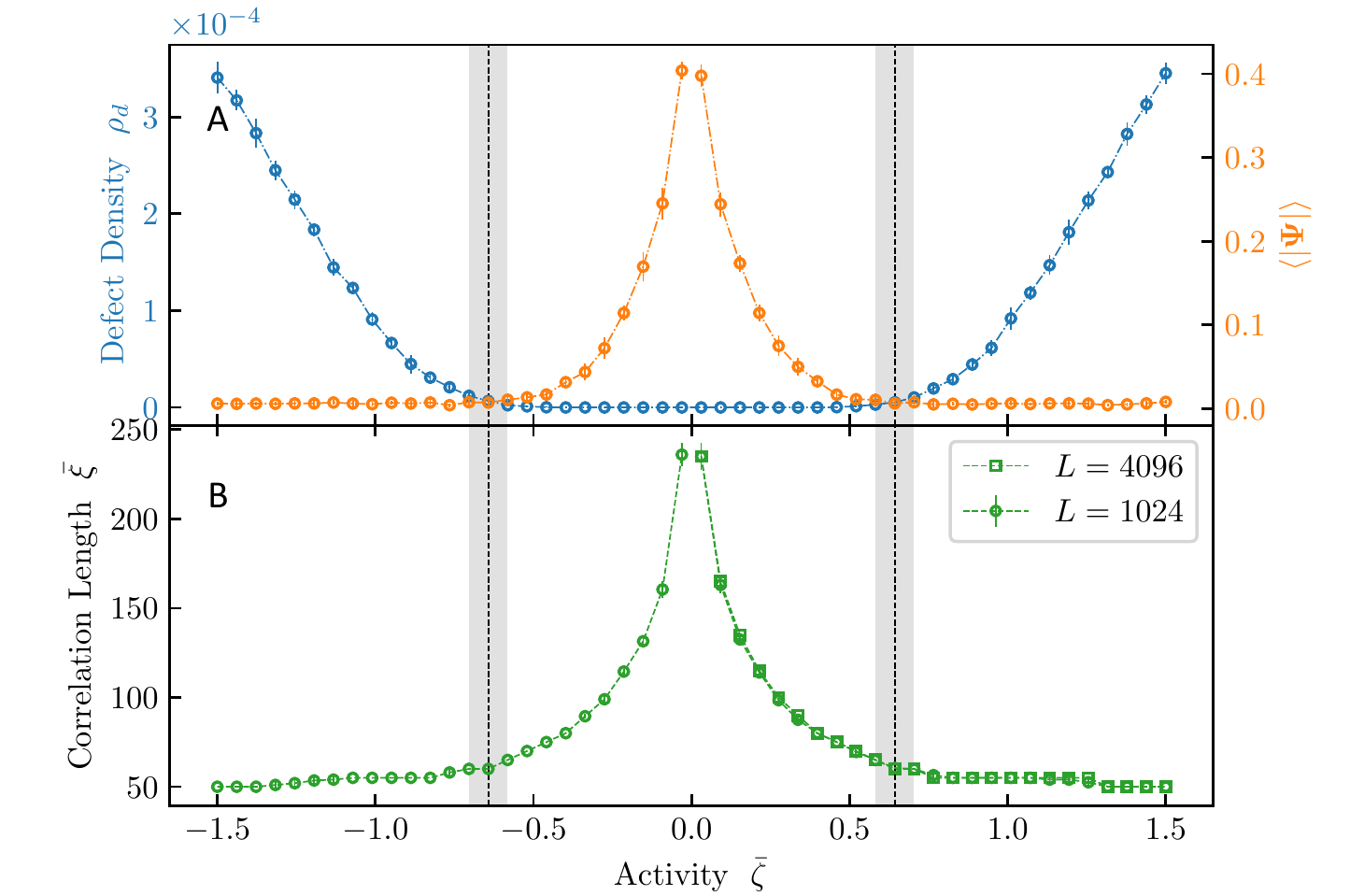}
    \caption{{\bf Global symmetry restoration.} (A) Averaged defect density $\rho_d$, {\it blue solid line}, and averaged magnitude of the global polarization $\mathbf{\Psi}$ (Eq.~\ref{eq:Psi}), {\it orange solid line}. (B) The correlation length $\bar{\xi}$ (in units of $l_p=\sqrt{K/A}$) defined by fitting an exponential to the pair-correlation function $\Gamma(r)$. The square and circle symbols in (B) correspond to the simulations results for $4096\times 4096$ and $1024\times 1024$ domains, respectively, showing no detectable difference.}
    \label{fig:correlationl}
\end{figure}

\subsection{Global symmetry is restored in the defect-laden active turbulence}
We next investigate whether the emergence of topological defects leads to alterations in the global ordering features of the polar active turbulence. To this end, we first introduce the global polarization $\mathbf{\Psi}$ as the spatiotemporal average value of the polarity field $\mathbf{p}$
\begin{equation}
    \mathbf{\Psi} = \int\frac{\mathrm{d}t}{T}\int\frac{\mathrm{d}^2\mathbf{x}}{L^2}\,\mathbf{p}(t, \mathbf{x}). \label{eq:Psi}
\end{equation}
The magnitude of this global polarization $|\mathbf{\Psi}|$ is a macroscopic order parameter for polar order akin to total magnetization in spin systems: $|\mathbf{\Psi}|=0$ is associated with disordered states (i.e. continuous rotational symmetry) while $|\mathbf{\Psi}|>0$ is associated with ordered states (discrete/broken rotational symmetry). We use this global measure to probe the crossover between the two active turbulence regime as shown in Fig.~\ref{fig:correlationl}A; {\it orange solid line}. Upon increasing the active stress from zero, the active turbulence state immediately manifests a broken $\SO(2)$-symmetry evident from a non-zero value of $|\mathbf{\Psi}|$ in the absence of topological defects. The global polarisation decreases in magnitude with increasing activity and, remarkably, drops to zero beyond a critical activity for the nucleation of topological defects. This is indicative of the restoration of the $\SO(2)$-symmetry in the active defect-laden turbulence. 

To further characterize the signature of this crossover, we also measure the equal-time, spatial pair-correlation function for the polarity field $\Gamma(\mathbf{r}) = \langle\delta p_i(\mathbf{x})\delta p_i(\mathbf{x} + \mathbf{r})\rangle$, where $\delta p_i(\mathbf{r}) = p_i(\mathbf{r}) - \langle p_i(\mathbf{r})\rangle$ is the deviation from the local mean value and 
the averaging is performed over $10$ different realizations, with distinct initial conditions and over time, once the system has reached a statistical steady state. Accordingly, the correlation length $\xi$ is obtained by fitting an exponential to the the pair-correlation function $\Gamma(r)$. Similar to the global polarization, the correlation length $\bar{\xi}=\xi/l_p$, normalized with respect to the coherence length $l_{p}=\sqrt{K/A}$, decreases with increasing activity, showing a fast decay that is followed by a significantly slower decay of the correlation length at higher activities (Fig.~\ref{fig:correlationl}). The regime shift is marked by the crossover from the fast to slow decay of the correlation length $\bar{\xi}$ which matched with the onset of defect nucleation and vanishing of $|\mathbf{\Psi}|=0$ as evidenced in 
Fig.~(\ref{fig:correlationl}A, B). We have carefully checked that correlation measurements are independent of the domain sizes for very large systems up to $4096 \times 4096$. We note that the behavior of the correlation length shares interesting similarity to that in frustrated two-dimensional Heisenberg magnets, where a crossover  between a topological defects dominated regime at high temperatures 
 and a spin-wave regime at low temperatures is observed~\cite{hasselmann2014interplay}.
Together, these measurements of the global polarization and the correlation length indicate that the crossover from defect-free to defect-laden active turbulence is marked by significant alterations in both the global symmetry of the collective organization and in the local correlations between the polarities of active constituents.

It is important to note that the observed restoration of the $\SO(2)$-symmetry is solely activity-driven and is in the noise-less limit of the equations of motion. Furthermore, the result presented in Fig.~\ref{fig:correlationl} demonstrate that, point-by-point, for any activity, (i) the correlation length, (ii) polarisation, and (iii) defect density for domain sizes $1024 \times 1024$ and $4096 \times 4096$ fall exactly on top of each other.
This strongly suggests a system-size independent behavior, at least for the numerically accessible system sizes. The question about the existence of a well-defined thermodynamic limit of this symmetry-breaking/restoring transition remains open for future studies. Additionally, it is constructive to note that we do not observe any difference between longitudinal and transverse correlations in the polarity. The correlation length does not diverge with the system size within the ordered state, which we conjecture is due to the non-equilibrium nature of the system and the breaking of detailed balance~\cite{tasaki2020hohenberg} through activity-induced flows that couple to the dynamics of the order parameter.
\begin{figure*}[t!]
    \centering
    \includegraphics[width=1.0\linewidth]{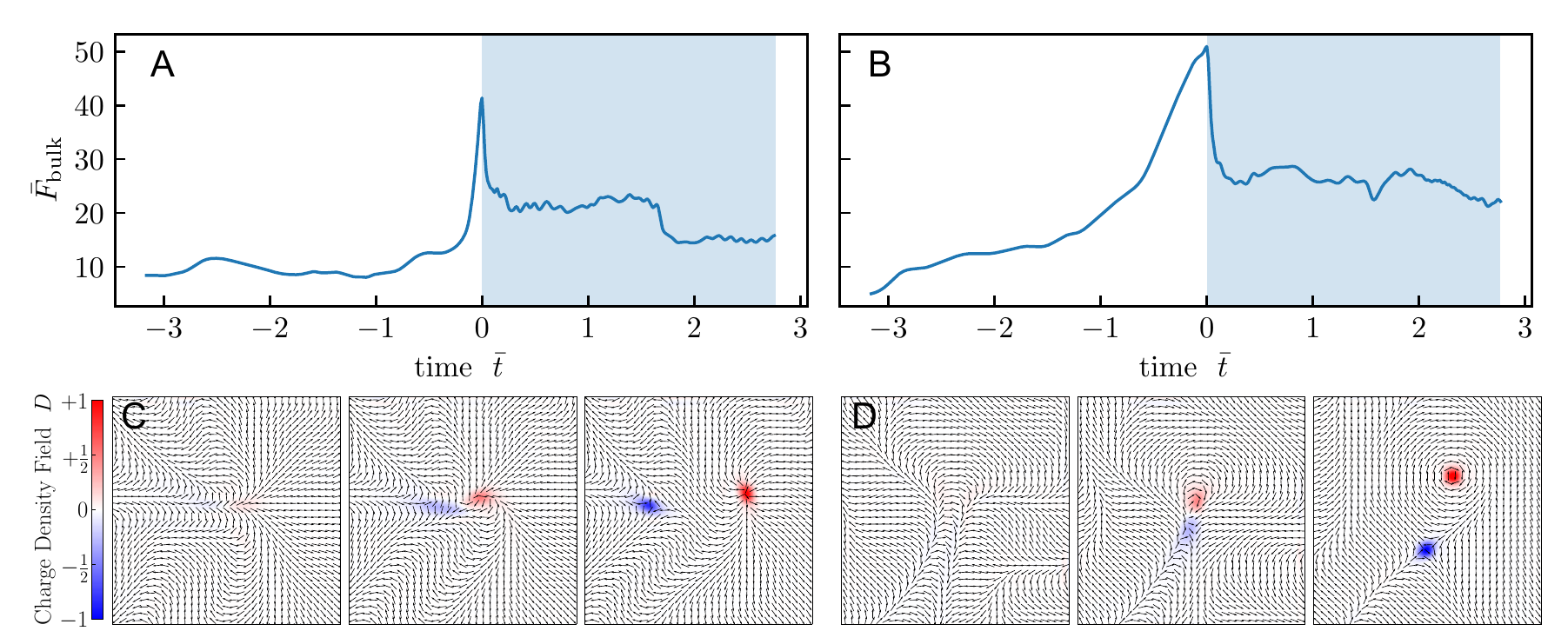}
    \caption{{\bf Defect pair nucleation process.} (A,B) Temporal evolution of the average bulk free-energy $\bar{F}_\mathrm{bulk} = (F_\mathrm{bulk} + A/4)/AL_\mathrm{act.}$ (top) measured in units of $AL_\mathrm{act.}$ relative to the background field $-A/4$; along with (C,D) simulation snapshots depicting the polarity field and diffuse charge density (heat plot) before, at, and after the nucleation of one pair of oppositely charged topological defects (bottom). Column (A,C) contractile and (B,D) extensile. The blue regions in A, B mark the period of existence of the defect pair considered. Time $\bar{t}$ is measured in units of the passive relaxation time of polarity over the simulation domain $t_p = \frac{\gamma L}{\sqrt{AK}}$ (see also Supplementary Movies~\cite{SI}).}
    \label{fig:freee_energy}
\end{figure*}
\subsection{Mechanism of defect pair nucleation in active turbulence of polar particles}
Having established the significant impact of the topological defects on the collective organization of polar active matter and its spontaneous flows, we next explore the mechanism of how topological defects are formed within the active turbulence state as the activity approaches the threshold value. Figure~\ref{fig:freee_energy} illustrates zoomed-in snapshots of the polarity field at the onset of one defect pair nucleation. Three snapshots are shown at simulation times corresponding to: before, at, and after a single pair nucleation for both extensile and contractile active stresses. To identify and track topological defects, we make use of the diffuse charge density, defined in the next section as $D = \frac{1}{2\pi}\epsilon_{ij}\epsilon_{kl}\partial_ip_k\partial_jp_l$: it carries the $\pm 1$ charge of the integer topological defect at the singularities in the polarity field, and is zero elsewhere. Shortly before the onset of defect pair nucleation, the polarity field is characterized by locally ordered domains separated by lines of kink walls (Fig.~\ref{fig:freee_energy} and Supplementary Movies~\cite{SI}). At the onset of defect pair nucleation the polarity within one of the kink walls flips locally, leading to spontaneous emergence of a pair of full-integer defects (Fig.~\ref{fig:freee_energy}). After the nucleation event, the defect pair gets separated by an ordered region of polarity alignment (Fig.~\ref{fig:freee_energy}). While the negatively charged antivortices ($-1$ topological defects) have similar structures in both extensile and contractile systems, the positively charged $+1$ topological defects take the form of asters and vortices in contractile and extensile systems, respectively (Fig.~\ref{fig:freee_energy}C,D), in agreement with earlier analytical predictions on the stability of defect structures in polar active matter~\cite{kruse2004asters}.

In order to gain more insight into the defect pair nucleation process, we measure the averaged bulk free energy of the system from Eq. (\ref{eq:F}) during the time leading to a single defect pair nucleation event, as depicted in (Fig.~\ref{fig:freee_energy}A-B). For both extensile and contractile systems the onset of defect nucleation coincides with a burst in the bulk free energy. This is consistent with the spontaneous flipping of the polarity within the kink walls that occurs when activity-induced fluctuations in the polarity overcome the energy barrier of local flipping, set by the bulk free energy. Once the defect pair is formed, the bulk free energy drops as the high energy stored in the kink walls is released. In the following section, we look more closely at this defect nucleation process and derive an analytical description of the activity threshold that is set by the competition between bulk and elastic energies of the system, and activity-induced flows.

\subsection{Onset of active defect-laden turbulence\label{sec:theory}}
To gain more theoretical insights into the onset of defect pair nucleation, we perform a stability analysis of the topological charge density field. A stability analysis of the polarity field to probe the nucleation of topological defects is unfeasible since it is intrinsically nonlinear. However, we take advantage of the property of the topological charge density field being zero for low-energy excitations of the polarity field (aka kinks, domain walls) to study the onset of defect nucleation as an instability to a nonzero and localised charge density field. To this end, we employ the Halperin-Mazenko method~\cite{halperin1981published,mazenko1997vortex,skogvoll2023unified} of
topological defects as zeros in the polarity vector field $\mathbf{p}$, i.e. localised regions where $\mathbf{p}$ vanishes in magnitude while its orientation is undefined (multi-valued). The associated topological charge is a quanta of the phase jump measured around an arbitrary contour $C_\alpha$ enclosing the defect 
\begin{equation}
2\pi q_\alpha = \oint\limits_{C_\alpha} \mathrm{d}\mathbf{l} \cdot \nabla \theta, 
\end{equation}
where $\theta = \arg(\mathbf{p})$. By Stokes' theorem, we see that topological defects correspond to phase singularities where the phase ceases to be irrotational,
\begin{equation}
2\pi q_\alpha \int\limits_{A_\alpha}\mathrm{d}^2\mathbf{x}\, \delta(\mathbf{x}-\mathbf{x}_\alpha) = \int\limits_{A_\alpha}\mathrm{d}^2\mathbf{x}\, \nabla\times \nabla\theta. 
\end{equation}
As shown in Ref.~\cite{skogvoll2023unified}, the topological content inside an area region $A_\alpha$ can be also obtained from the corresponding integration in the order parameter space $(p_1,p_2)$  
\begin{equation}
2\pi q_\alpha \int\limits_{A_\alpha}\mathrm{d}^2\mathbf{x}\, \delta(\mathbf{x}-\mathbf{x}_\alpha) = \int\mathrm{d}^2\mathbf{p}  \mathrm{det}(\nabla\mathbf p) \delta(\mathbf{p}). 
\end{equation}
We relate the topological defect density $D$ with the Jacobi determinant of the transformation to the order parameter space, which is the same as the determinant of the distortion tensor $\nabla\mathbf{p}$ to define
\begin{equation}
    D = \frac{1}{2}\epsilon_{ij}\epsilon_{kl}\partial_ip_k\partial_jp_l,
\end{equation}
which is a smooth scalar field that vanishes where $\mathbf{p}$ is smooth and becomes non-zero and localized at the core topological defect picking up the phase singularity. We have validated the defects detected by D-filed against our routine approach of using winding angles and have obtained identical results, in agreement with previous studies on active nematics, where the $D$-field was defined as the diffuse charge density~\cite{blow2014biphasic,doostmohammadi2016defect,saw2017topological}. The $D$-field has the physical interpretation of a non-singular topological charge density field which follows the conservation law ~\cite{skogvoll2023unified}
\begin{equation}\label{eq:evolution_D}
    \partial_t D + \partial_iJ_i =0,
\end{equation}
with the corresponding topological current density
\begin{equation}
    J_i = -\epsilon_{ij}\epsilon_{kl}\partial_tp_k\partial_jp_l,
\end{equation}
which is fully determined by the evolution of the polarity $\mathbf{p}$ and the flow field $\mathbf u$ through the main Eqs.~(\ref{eq:p}-\ref{eq:u}).

We perform a stability analysis of Eq.~(\ref{eq:evolution_D}) around the ground state of zero $D$ associated with slowly-varying polar order to estimate the critical activity for the onset of defect nucleation, identified as an instability where $D$ becomes non-zero because polar order vanishes locally. In general, the evolution of $D$, resulting from inserting the evolution of $\mathbf p$ from Eq.~(\ref{eq:p}) into Eq.~(\ref{eq:evolution_D}),  cannot be closed only in terms of the $D$-field. However, in the limit of retaining only the shear flow alignment contribution and the molecular term, the evolution of $\mathbf p$ from Eq.~(\ref{eq:evolution_D}) can be reduced to  
\begin{equation}\label{eq:evolution_D_closed}
    \partial_t D = \epsilon_{ij}\epsilon_{kl}\left[\lambda\partial_i(E_{kn}p_n)\partial_j p_l+\frac{1}{\gamma}\partial_i h_k\partial_j p_l\right].
\end{equation}

Now, by balancing viscous stress with the active stress in Eq.~(\ref{eq:u}), we relate the traceless strain rate $E_{kn}$ directly to the nematic order parameter, namely
\begin{equation}\label{eq:E_Q_rel}
    E_{kn} = \frac{\zeta}{2\eta} \left(p_np_k-\frac{p^2}{2}\delta_{nk}\right),
\end{equation} 
such that the flow alignment contribution dominated by orientational distortion has a closed form given by 
\begin{eqnarray}
&\frac{\zeta\lambda}{4\eta}\epsilon_{ij}\epsilon_{kl}\partial_i(p^2p_k)\partial_j p_l = \nonumber \\
&\frac{\zeta\lambda}{4\eta}\left(
p^2\epsilon_{ij}\epsilon_{kl}\partial_ip_k\partial_j  p_l + 2\epsilon_{ij}\epsilon_{kl} p_k p_m\partial_ip_m \partial_j  p_l\right) = \nonumber
\\ 
&\frac{\zeta\lambda}{2\eta} p^2 D + \frac{\zeta\lambda}{2\eta}p^2 (\partial_1 p_1 \partial_2 p_2 - \partial_2 p_1 \partial_1 p_2 ) = 
\frac{\zeta\lambda}{\eta}p^2 D. 
\end{eqnarray}
The molecular field is due to the elastic distortions and the ordering potential
\[h_k = K\nabla^2 p_k +A (1 -p^2)p_k,\]
and leads to the contribution
\[\epsilon_{ij}\epsilon_{kl}\partial_ih_k\partial_j p_l= K \nabla^2 D+ 2A(1-2p^2) D.\]
Thus, within this limiting case, the evolution of the $D$-field reduces to a diffusion-reaction equation 
\begin{equation}\label{eq:evolution_D_approx}
\partial_t D \approx \frac{K}{\gamma}\nabla^2  D+ \left(\frac{2A}{\gamma}(1-2p^2) +\frac{\zeta \lambda}{\eta}p^2\right)D. 
\end{equation}

First, we notice that the polarity stiffness $K$ enters as en effective diffusivity coefficient and sets the scale of the diffusive core. The kinetic rate coefficients arise from the stabilizing passive contribution through the energy scale parameter $A$ of the local order potential and its relaxation timescale $\gamma$, as well as from the destabilizing contribution through the flow alignment $\lambda$, activity $\zeta$ and viscosity $\eta$.
At the first sight, Eq.~(\ref{eq:evolution_D_approx}) looks non-conservative due to the source/sink term. 
This is, however, not the case since this term is derived from a conservative current and is related to the fact that the $D$-field itself is the divergence of a vector field, $D =\partial_i(\epsilon_{ij}\epsilon_{kl}p_k\partial_jp_l)/2$.
We estimate the critical activity from the linear stability analysis of Eq.~(\ref{eq:evolution_D_approx}) about the ordered state $p=1+\delta p$ and $D=\delta D$.
By balancing out the two contribution to the reaction term we find that the equation becomes unstable at 
\begin{equation}
    \zeta_c\approx  \frac{2 \eta}{\gamma\lambda} A.
\end{equation}
Thus, in the limit where we neglect the convective and rotational terms to the evolution of the $D$-field (i.e. $\mathbf u\cdot \nabla \mathbf p$ and $[\mathbf \Omega,\mathbf p]$), the topological charge density becomes unstable to defect nucleation only for extensile systems where $\zeta>0$. This instability is due to the flow alignment. 

Advection and rotation of polarity $\mathbf p$, due to flow velocity and vorticity, could also trigger instabilities in the $D$-field. Alas, their contributions are non-locally through viscosity and incompressibility constraint and cannot be reduced to linear operators acting on $D$. However, it turns out that the contribution of vorticity to the $D$-evolution is symmetric with respect to the sign of $\zeta$, and, from this, we can infer that an instability induced by vorticity would occur at a critical activity both for extensile/contractile systems, also symmetric with respect to the sign of $\zeta$.

It is important to note that in numerical simulations all these destabilizing forces orchestrate and compete with the relaxation to uniform order. Therefore, the value of the critical activity is different than the estimated one corresponding to an isolated triggering factor. However, this stability analysis is informative and predicts generic properties of $\zeta_c$: i) it depend on the energy scale $A$ controlling polar order, and ii) that the threshold activity has an asymmetry with respect to extensile/contractile active stresses in the presence of flow alignment.

\begin{figure}[t!]
    \centering
    \includegraphics[width=1.0\linewidth]{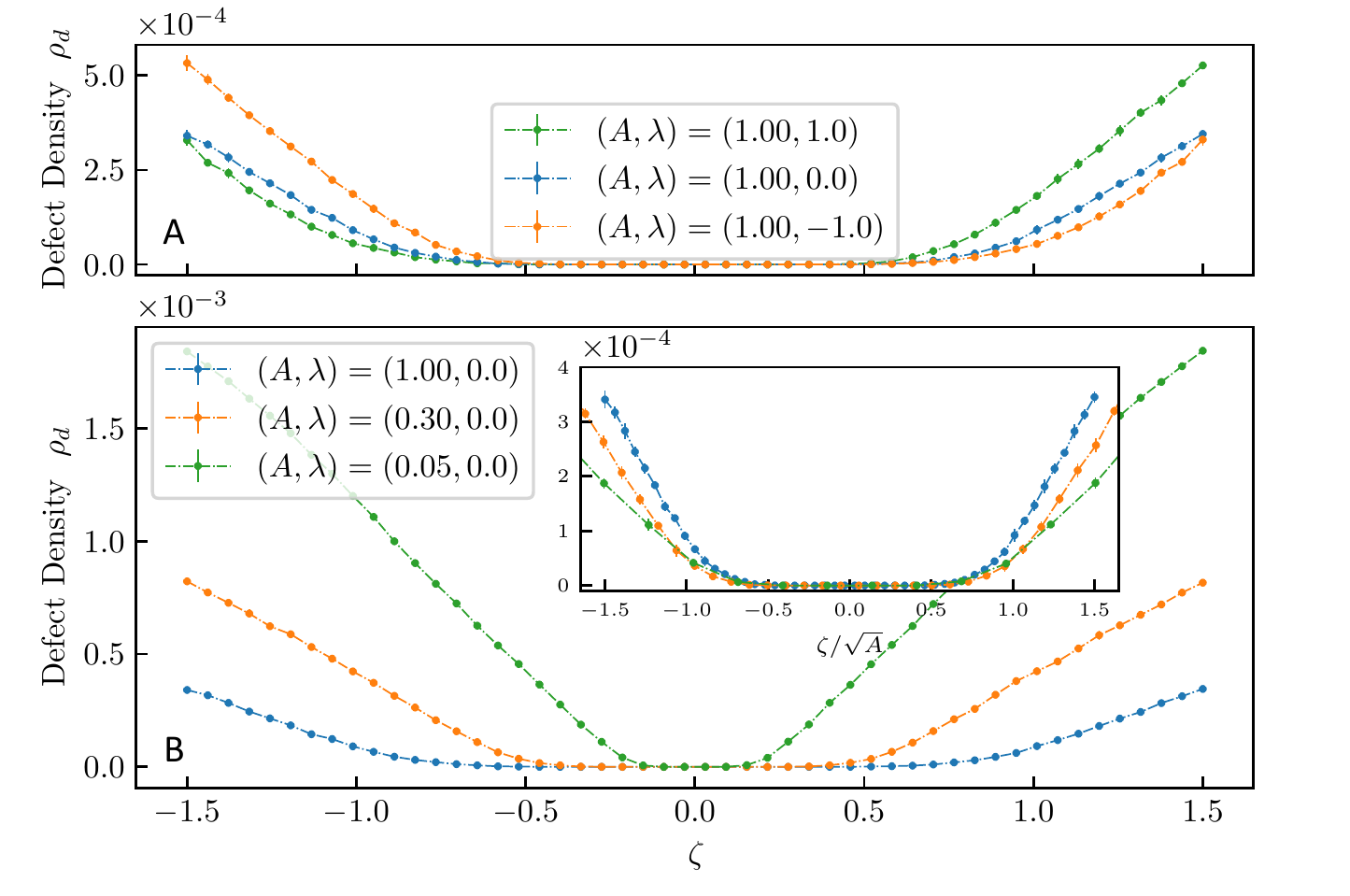}
    \caption{Dependence of the steady-state density of defects on (A) the flow-aligning parameters $\lambda$ and (B) the bulk free-energy strength A. Inset in (B) shows that normalizing the active stress by $\sqrt{A}$ results in a data collapse near the crossover point.}
    \label{fig:number_of_defects}
\end{figure}

To test the theoretical predictions on the activity threshold for defect nucleation, we next explore the dependence of defect density on the flow-aligning parameter $\lambda$ and the energy scale $A$ for the bulk free energy. Figure~\ref{fig:number_of_defects}A compares the variation of defect density $\rho_d$ with activity for three values of the flow-aligning parameter $\lambda$. Interestingly, and consistently with the theoretical prediction, for $\lambda=0$ the activity threshold for defect nucleation is symmetric with respect to the sign of the activity. Additionally, as predicted from the theory, a positive (negative) flow-aligning parameter shifts the defect density curve such that the activity threshold becomes smaller - in absolute value - for extensile (contractile) activities compared to their contractile (extensile) counterparts, leading to a more asymmetric profile of the defect density with respect to the sign of the activity. Moreover, as predicted from the theory, decreasing the energy scale in the bulk free energy $A$ lowers the activity threshold for defect nucleation (Fig.~\ref{fig:number_of_defects}B). This is also consistent with the mechanism identified for the nucleation of pairs of defects, relying on the spontaneous flipping of the polarity within the kink walls (Fig.~\ref{fig:freee_energy}C,D and Supplementary Movies~\cite{SI}): since the parameter $A$ sets the depth of a double-well potential in the bulk free energy, decreasing its value amounts to lowering the energy barrier for flipping the polarity, and as such leads to a reduced activity threshold for defect pair nucleation.
\subsection{Universal scaling in defect-free and defect-laden active turbulence}
\begin{figure*}[t!]
    \centering
    \includegraphics[width=1.0\linewidth]{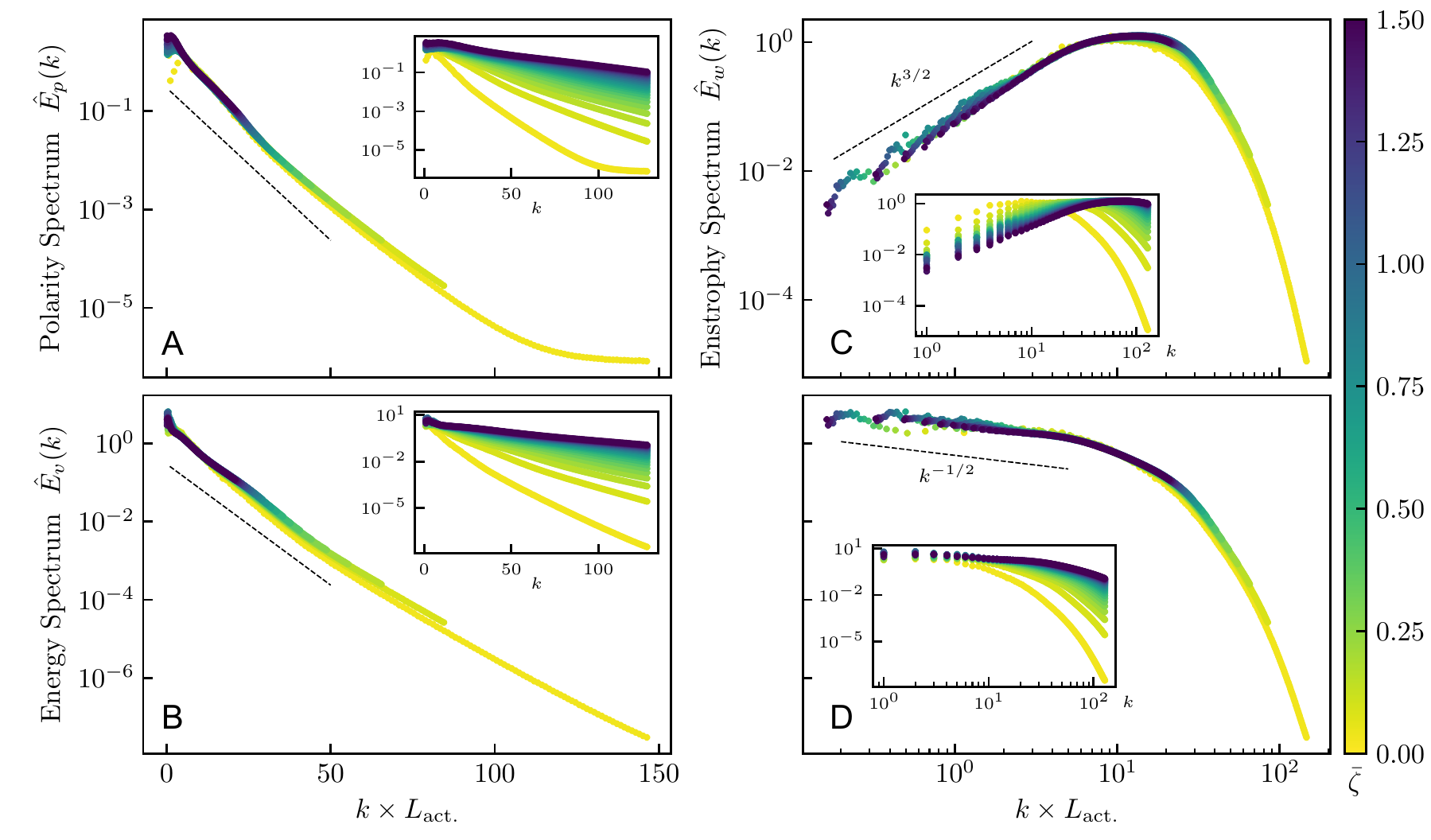}
    \caption{{\bf Universal scaling in defect-free and defect-laden active turbulence.} Power spectra normalized with its corresponding maximum value for (A) polarity $\hat{E}_p(k)$ and (B) kinetic energy $\hat{E}_v(k)$ on semilog plots, demarcating the exponential decay at small scales (large wave numbers) that imply the dynamics is dominated by viscous dissipation.
    (C,D) Power spectra normalized with its corresponding maximum value for (C) the enstrophy $\hat{E}_w(k)$ and (D) kinetic energy again, on log-log plots, showing the power-law scaling at scales larger than the active length scale $L_\mathrm{act.}$. Insets show the non-scaled power spectra and the colorbar indicates the dimensionless active stress strength $\bar{\zeta}$. }
    \label{fig:spectra}
\end{figure*}

Having established the mechanism of defect nucleation and the difference between defect-free and defect-laden active turbulence, we now investigate the common features between these two active turbulent states by characterizing the energetic features of the flow and order parameter (Fig.~\ref{fig:spectra}). To this end, we first measure the power spectrum of the polarity order parameter $\hat{E}_{p}(k) = \tfrac{1}{2}\langle \hat{p}_i(k) \hat{p}_i(-k) \rangle$, with $k$ the wave number averaged over azimuthal direction, for a range of activities spanning both defect-free and defect-laden turbulent states. Interestingly, when the wave number is non-dimensionalized by the active length scale $L_{\text{act.}}=\sqrt{K/|\zeta|}$, which characterizes the scale of energy injection into the system by active stresses, the power spectrum $\hat{E}_p$ for all activity values shows an exponential decay at {\it wave numbers larger than the inverse active length scale}, as evident from a semilog plot of the order parameter spectrum versus the normalized wave number (Fig.~\ref{fig:spectra}A). Notice that the exponential cutoff scale still depends on activity and approaches a constant value at sufficiently high activity. This is related to the initial fast decay of the correlation function $\Gamma(r)$, with a correlation length that is activity dependent as discussed earlier. Notice that the exponential dependence of $\hat{E}_p$ on $k$ breaks
down for small wavenumbers. On the other hand, robust exponential tails at large wave numbers are characteristic for the spectral properties of the spontaneous flow, measured by the kinetic energy spectrum $\hat{E}_{v}(k) = \tfrac{1}{2}\langle \hat{v}_i(k) \hat{v}_i(-k) \rangle$, with $\hat v_i$ the Fourier transforms of flow velocity (Fig.~\ref{fig:spectra}B). However, for {\it wave numbers below the inverse active length scale}, the kinetic energy spectrum $\hat{E}_{v}(k)$ and enstrophy spectrum $\hat{E}_{\omega}(k) = \tfrac{1}{2}\langle \hat{\omega}(k) \hat{\omega}(-k) \rangle\sim k^{2} \hat E_v(p)$, both exhibit power-law behavior (Fig.~\ref{fig:spectra}C,D).
This is important, because these activity levels span both the defect-free and defect-laden active turbulence states, indicating that for a polar active matter, despite distinct symmetry features and correlation lengths, there exist universal scaling of flow and order parameter spectrum in active turbulence state with and without topological defects. In particular, the enstrophy power spectrum that relates to the flow field shows a universal {\it power-law} scaling at length scales larger than the active length $L_{\text{act.}}$, while the polarity power spectrum that relates to the order parameter field shows a universal {\it exponential decay} at length scales smaller than the active length scale. These two distinct scaling relate defect-free and defect-laden turbulence in polar active matter.
It is important to note that the scaling observed here is different from the power-law scaling suggested for active nematic turbulence without topological defects, where the orientational field is treated as a unit vector field with fixed modulus~\cite{alert2020universal}. The heuristic argument of the scaling behavior reported in Ref.~\cite{alert2020universal} is based on a spectral analysis of the equation for the vorticity source to argue for the power-law scaling of $\hat E_p(k)$ and $\hat E_v(k)$ at small $k$. The scaling argument hinges on two important assumptions: 1) vorticity source is dominated by the active stress determined only by phase gradients; 2) at small $k$ (long wavelength) phase fluctuations become uncorrelated. Both of these assumptions fail to apply when the polarity magnitude is allowed to vary, since the vorticity fluctuations couple with fluctuations in both magnitude and phase of the polarity. Thus, the power spectrum of vorticity source is not trivially related to polarity power spectrum by extending the scaling analysis from Ref.~\cite{alert2020universal}. Therefore, we do not expect to observe the same scaling behavior at long-wavelength as in Ref.~\cite{alert2020universal}.

\section{Discussion}
The results presented herein reveal distinct flow fields and patterns of collective self-organization of polar active particles that are primarily controlled by active stresses.
We show that increasing active stress in a polar active fluid leads to a dynamical crossovers from a defect-free to a defect-laden active turbulence state. Above a critical activity threshold, the turbulent flow is seeded with topological defects nucleated in the polar order parameter. Importantly, we find that the proliferation of defects screens the global polar order. Thus, in the defect-laden turbulent states the $\SO(2)$-symmetry is restored, whereas the defect-free active turbulence is endowed with broken $\SO(2)$-symmetry. 

Interestingly, we show that in spite of their distinct topological content, the two active turbulence regimes share similar statistical properties with robust exponential tails for power spectra of polar order, flow velocity and vorticity on length scales smaller than the active length scale. On the other hand, over the length scales that are larger than the active length scale, enstrophy and energy spectra of flow fluctuations exhibit self-similarity with power-law scaling exponents. These statistical properties are also universal in the sense that they do not depend on the presence of topological defects, which may indicate that non-topological dissipative structures (e.g. kink walls) are the dominant driving force to fluctuations in order and flow of polar active fluids. 

It is noteworthy that despite the ubiquitous presence of polarity in living and synthetic active materials~\cite{Marchetti13,Bechinger2016}, majority of the current understanding of topological defects and flow features within the active turbulence is based on studies of active nematics. This is in part due to the growing number of biological systems that are identified with half-integer, nematic, topological defects~\cite{doostmohammadi2021physics,shankar2022topological,ardavseva2022topological} and discovery of potential biological functions for such defects~\cite{Saw17,Kawaguchi17,copenhagen2020topological}. It is important to note, however, that full-integer defects have also been identified as possible organization centers for mitotic spindles in microtubule-motor protein assemblies~\cite{roostalu2018determinants} and more recently as potential sites for mechanotransduction~\cite{endresen2019topological}, and for cell differentiation in mouse myoblasts~\cite{guillamat2022integer} and cartilage cells~\cite{makhija2022topological}. Our results provide a quantitative characterisation of the dynamical crossovers, as well as the statistical imprints of flow and topological defects in polar active matter. Future work should focus on details of the dynamic distinctions in topological and flow features between the active turbulence in polar and nematic active materials, as well as generalized frameworks of the coupling between polar and nematic symmetries that have already been suggested theoretically~\cite{baskaran2010nonequilibrium,patelli2019understanding,Amiri_2022} and observed experimentally~\cite{roostalu2018determinants,makhija2022topological}.
\section*{Acknowledgement and Funding}
A. D. acknowledges funding from the Novo Nordisk Foundation (grant No. NNF18SA0035142 and NERD grant No. NNF21OC0068687), Villum Fonden Grant no. 29476, and the European Union via the ERC-Starting Grant PhysCoMeT. J.R. and L.A. acknowledge support from the Research Council of Norway through the Center of Excellence funding scheme, Project No. 262644 (PoreLab). 
\bibliographystyle{unsrt}
\bibliography{references}
\end{document}